\documentclass{ws-mpla}

 \usepackage{hyperref}
 \usepackage[super]{cite}
 \usepackage{mathrsfs}
 \usepackage{xcolor}
 \usepackage[utf8]{inputenc}
 \usepackage{upgreek}
 \usepackage{amsmath,amssymb}
 \usepackage[english]{babel}
 \usepackage{wrapfig}
 \usepackage{graphics}
 \usepackage{graphicx}
 \usepackage{epstopdf}
 \usepackage{url} 
 
\renewcommand{\vec}{\boldsymbol}
\newcommand{\wrho}{\widetilde{\rho}}
\newcommand{\st}{{\,\star}}

\def\URL#1{\url{<#1>}} 

\begin{document}

\markboth{V.~A.~Naumov and D.~S.~Shkirmanov}
{Covariant asymmetric wave packet for a field-theoretical description of neutrino oscillations}

\catchline{}{}{}{}{}

\title{Covariant asymmetric wave packet \\ for a field-theoretical description of neutrino oscillations}

\author{\footnotesize VADIM~A.~NAUMOV           
                  and DMITRY~S.~SHKIRMANOV      
       }
\address{Bogoliubov Laboratory of Theoretical Physics, Joint Institute for Nuclear Research, \\ 
RU-141980 Dubna, Russia, vnaumov@theor.jinr.ru, shkrimanov@theor.jinr.ru}

\maketitle

\pub{Received (Day Month Year)}{Revised (Day Month Year)}

\allowdisplaybreaks

\begin{abstract}
We consider a class of models for the relativistic covariant wave packets which can be used as asymptotically free
in and out states in the quantum field theoretical formalisms for description of the neutrino flavor oscillation phenomenon.
We demonstrate that the new ``asymmetric'' wave packet (AWP) is an appropriate alternative to the more
conventional ``symmetric'' wave packets, like the so-called relativistic Gaussian packet (RGP) widely used in
the QFT-based approaches to neutrino oscillations. We show that RGP is not a particular case of AWP,
although many properties of these models are almost identical in the quasistable regime.
We discuss some features of AWP distinguishing it from RGP. 
\end{abstract}

\ccode{PACS Nos.: 03.70.+k, 14.60.Pq, 14.60.Lm}


\section{Introduction}

There are several approaches to description of the neutrino flavor oscillations in vacuum and matter, 
started from the conventional quantum-mechanical (QM) approach\cite{Gribov:1968kq,Bilenky:1978nj}
and developed into the more advanced approaches based on the methods of relativistic quantum field theory (QFT).
\cite{Giunti:1993se,Grimus:1996av,Mohanty:1997zi,Giunti:1997sk,Campagne:1997fu,Kiers:1997pe,Zralek:1998rp,Ioannisian:1998ch,%
      Grimus:1998uh,Grimus:1999zp,Grimus:1999ra,Cardall:1999bz,Cardall:1999ze,Stockinger:2000sk,Beuthe:2000er,Beuthe:2001rc,%
      Chang:2001mh,Beuthe:2002ej,Giunti:2002xg,Garbutt:2003ih,Asahara:2004mh,Fujii:2004px,Dolgov:2005nb,Nishi:2005dc,Akhmedov:2008jn,%
      Naumov:2009zza,Kopp:2009fa,Akhmedov:2009rb,Keister:2009qn,Bernardini:2010zba,Anastopoulos:2010sf,Naumov:2010um,%
      Akhmedov:2010ua,Akhmedov:2010ms,Morris:2011sn,Akhmedov:2012uu,Akhmedov:2012mk,Blasone:2014jea,Law:2014tva}
Despite the fact that the standard QM theory is commonly used for interpretation of the experimental
data and extraction of the neutrino mixing parameters, it has a lot of internal inconsistencies
(see, e.g., Ref.~\refcite{Giunti:2007ry} and references therein). This particularly stimulated the
development of the QFT-based approaches that are free of the most of these inconsistencies. 
These new approaches determine the domain of applicability of the QM theory and, as a bonus, predict
potentially measurable deviations from the standard formulas for the flavor transition probabilities
(see, e.g., Refs.~\refcite{Stockinger:2000sk,Naumov:2013bea}).
In this paper, we explore the particular QFT formalism in which the process of neutrino production
and detection is described by the so-called macroscopic Feynman diagrams
and the neutrino flavor oscillation phenomenon is nothing but a result of interference of the diagrams
perturbatively describing the lepton number violating processes with the neutrino mass eigenfields
as internal lines (propagators) connecting the production and detection vertices of the diagram
(``source'' and ``detector'', respectively).
The external lines of a macrodiagram (in and out one-particle states) are described by asymptotically free
quasistable wave packets (WP) constructed as covariant superpositions of the standard one-particle Fock states
$|\vec{p},s,\ldots\rangle$ with definite momenta $\vec{p}$, spin projection $s$, etc. These superpositions
must satisfy some natural conditions, including the correspondence principle, according to which WP state
reduces to the Fock state $|\vec{p},s,\ldots\rangle$ in the plane-wave limit.\cite{Naumov:2009zza,Naumov:2010um}
The correspondence principle is one of the key constraints which allows to determine the most important properties
of the admissible WP states, but it is, of course, not sufficient for an unambiguous determination of all of them.

Some relevant features of the QFT WP states were studied 
recently\cite{Takeuchi:1999as,Matsuo:2005fb,Korenblit:2013-BSFF2013,Naumov:2013vea,Perez:2013uca,Karlovets:2014wva}
but the goal of the present paper is different from that in the cited studies. 
To elucidate our reasons we note that even though the neutrino mass eigenfield is pure virtual within
the diagrammatic approach, it turns out that effectively it can be treated as a real on-mass shell particle.
In other words, it can be treated as an effective wave packet.
In an asymptotic regime, when the spatial distance, $L$, between the source and detector vertices is sufficiently
large, this effective neutrino wave packet has the same form as the external in and out states
if the latter are described by the so-called relativistic Gaussian packets (RGP). \cite{Naumov:2010um}
If however the distance $L$ is short, the neutrino wave packet acquires the form quite different from
the RGP one.\cite{Alushta-2012} In particular, it is not spherically symmetric, even in the intrinsic frame
of reference of the (massive) neutrino. It signifies that the RGP model may be inadequate in some circumstances. 
Moreover, some features of the external wave packets may affect the differences between the QM and QFT
predictions, such as potentially measurable decoherence effects.
It is therefore important to study the model-dependent properties of the WP states. 

\section{Generic wave-packet states in QFT.}
\label{Wave-packetStates}

Let us start with a heuristic exploration of the relativistic wave packet (WP) suitable for description of
a ``particle-like'' quantum state.
For simplicity, we first consider the quantum-mechanical picture and neglect the spin variable.
Let $|\vec{k}\rangle$ be the eigenstate of the on-shell (with mass $m$) 4-momentum operator
$\hat{P}=(\hat{P}_0,\hat{\vec{P}})$. 
Hereafter, we denote $k_0=E_{\vec{k}}=\sqrt{\vec{k}^2+m^2}$ and use the following normalization:
$\langle\vec{k}'|\vec{k}\rangle = (2\pi)^32E_{\vec{k}}\delta(\vec{k}'-\vec{k})$. 
By definition, ${\hat{P}_\mu|\vec{k}\rangle=k_\mu|\vec{k}\rangle}$ ($\mu=0,1,2,3$) and thus
$\hat{P}^2|\vec{k}\rangle=m^2|\vec{k}\rangle$.
Consider now an abstract ``one-particle'' spinless state $|a\rangle$.
It can be decomposed into the 3-momentum basis $\{|\vec{k}\rangle\}$ that is be represented as a 
packet of plane waves:
\begin{equation*}
\label{a_k}
{|a\rangle = \int \frac{d\vec{k}\,\uppsi_{\vec{k}}}{(2\pi)^3\sqrt{2E_{\vec{k}}}}|\vec{k}\rangle},
\quad
{\uppsi_{\vec{k}}=\frac{\langle\vec{k}|a\rangle}{\sqrt{2E_{\vec{k}}}}}.
\end{equation*}
Similarly, the state $|a\rangle$ can be decomposed into the eigenvectors $|\vec{x}\rangle$ of the position operator
$\hat{\vec{X}}=(\hat{X}_1,\hat{X}_2,\hat{X}_3)$ defined by $\hat{X}_i|\vec{x}\rangle=x_i|\vec{x}\rangle$ ($i=1,2,3$)
and normalized as $\langle\vec{x'}|\vec{x}\rangle=\delta\left(\vec{x'}-\vec{x}\right)$; we obtain:
\begin{equation*}
\label{a_x}
{|a\rangle = \int d\vec{x}\,\uppsi_{\vec{x}}|\vec{x}\rangle},
\quad
{\uppsi_{\vec{x}}=\langle\vec{x}|a\rangle}.
\end{equation*}
Clearly, the wave functions $\uppsi_{\vec{k}}$ and $\uppsi_{\vec{x}}$ are the Fourier transforms of each other:
\begin{equation*}
\label{chi_k-chi_x}
{\uppsi_{\vec{x}} = \int\frac{d\vec{k}}{(2\pi)^3}e^{i\vec{k}\vec{x}}\,\uppsi_{\vec{k}}},
\qquad
{\uppsi_{\vec{k}} = \int d\vec{x} e^{-i\vec{k}\vec{x}}\,\uppsi_{\vec{x}}}.
\end{equation*}

Let us take up first an \emph{unphysical} limiting case when the state $|a\rangle$ is localized in a point of space,
say $\vec{x}_a$, that is $|a\rangle=\text{const}\,|\vec{x}_a\rangle$.
Then $\uppsi_{\vec{x}}=\text{const}\,\delta(\vec{x}-\vec{x}_a)$ and $\uppsi_{\vec{k}}=\text{const}\,e^{-i\vec{k}\vec{x}_a}$.
Of course, such a state cannot be assigned to a real physical particle, since its momentum is completely uncertain.
Moreover, the particle cannot be localized in a region smaller than its Compton length $\sim1/m$.
It is however important that, in this \emph{mathematical} limit, both the wave functions
$\uppsi_{\vec{x}}$ and $\uppsi_{\vec{k}}$ depend explicitly on the spatial coordinate $\vec{x}_a$.
In the real world, any physical (particle-like) state $|a\rangle$ is localized within a finite spatial region $S_a$
or, more precisely, the probability density $|\uppsi_{\vec{x}}|^2$ quickly vanishes far from the domain $S_a$.
In general, $S_a$ can be described by a set of equations, inequalities, or coordinates.
We will limit ourselves to the simplest case, when the domain $S_a$ can be characterized by a single
3-vector $\vec{x}_a$, the center of its symmetry (e.g.\ $S_a$ is a sphere with the center in $\vec{x}_a$).
Then both $\uppsi_{\vec{x}}$ and $\uppsi_{\vec{k}}$ must be functions of $\vec{x}_a$.

Similarly, if the state $|a\rangle$ has a finite life-time, the wave functions $\uppsi_{\vec{x}}$ and $\uppsi_{\vec{k}}$
must be functions of a time variable $x_a^0$.
In the more general case of a space-time localization, the wave functions depend on the variables $\vec{x}_a$ and $x_a^0$.
Since however any Lorentz boost entangles the space and time variables, the wave functions $\uppsi_{\vec{x}}$ and $\uppsi_{\vec{k}}$
must depend on the 4-vector $x_a=(x_a^0,\vec{x}_a)$ which describes the localization of the state $|a\rangle$ in the Minkowski space-time.

In a similar spirit we may consider the  localization in the momentum space (plane wave limit).
Namely, we assume that the state $|a\rangle$ has a definite 3-momentum $\vec{p}_a$ that is $|a\rangle=\text{const}\,|\vec{p}_a\rangle$.
Then $\uppsi_{\vec{k}}=\text{const}\,(2\pi)^3\sqrt{2E_{\vec{k}}}\delta(\vec{k}-\vec{p}_a)$ and
$\uppsi_{\vec{x}}=\text{const}\sqrt{2p_a^0}e^{i\vec{p}_a\vec{x}}$.
Of course, this state is also unphysical since it is fully delocalized in the Minkowski space-time.
In the real world, any physical (particle-like) state $|a\rangle$ is localized within some finite region $M_a$ 
of the momentum space, in the sense that the probability density $|\uppsi_{\vec{k}}|^2$ quickly vanishes far
from the domain $M_a$. The arguments similar to the above allow us to conclude that both $\uppsi_{\vec{x}}$ and
$\uppsi_{\vec{k}}$ must be functions of $\vec{p}_a$.
Needless to say that the energy variable $p_a^0$ (in contrast with the time variable ${x_a^0}$) is not independent.

Finally we may conclude that the simplest WP state ${|a\rangle}$ suitable for description of a quantum particle
localized in both the configuration space and momentum space must depend on the space-time variable
$x_a$ and momentum variable $\vec{p}_a$, that is
$\uppsi_{\vec{k}}=\uppsi_{\vec{k}}(\vec{p}_a,x_a)$, $\uppsi_{\vec{x}}=\uppsi_{\vec{x}}(\vec{p}_a,x_a)$, and
$|a\rangle=|\vec{p}_a,x_a\rangle$.

Going to the field-theoretical picture, we may assume, by analogy with the QM case, that the relativistic WP state
can be characterized by an on-shell 4-momentum $p=(p_0,\vec{p})$, space-time variable $x=(x_0,\vec{x})$, and spin projection $s$.
In the momentum representation, we can build the WP state as a linear superposition of the Fock states~\cite{Naumov:2010um}.
Then the most general construction of the WP state is
\begin{equation}
\label{WPState}
|\vec{p},x,s\rangle = \int\frac{d\vec{k}}{(2\pi)^32E_{\vec{k}}}
\sum_{s'}\Upphi_{s's}\left(\vec{k},\vec{p},x;{{\vec{\sigma}}}\right)|\vec{k},s'\rangle,
\end{equation}
where $|\vec{k},s\rangle=\sqrt{2E_{\vec{k}}}a_{\vec{k}s}^\dag|0\rangle$
is the usual Fock one-particle state with definite momentum $\vec{k}$, energy $E_{\vec{k}}=k_0=\sqrt{\vec{k}^2+m^2}$
($m$ is the mass of the particle) and spin projection $s$; $\vec{\sigma}=\{\sigma_1,\sigma_2,\ldots\}$
is a set (finite or infinite) of parameters (constants or Lorentz scalars) governing the shape of
the WP state in the momentum space. 
The conventional (anti)commutation relations for the creation/annihilation operators hold:
\[
\{a_{\vec{q}r},a_{\vec{k}s}\}=\{a_{\vec{q}r}^\dagger,a_{\vec{k}s}^\dagger\}=0,
\quad
\{a_{\vec{q}r},a_{\vec{k}s}^\dagger\}=(2\pi)^3\delta_{s's}\delta\left(\vec{k}-\vec{q}\right).
\]

The proper Lorentz transformation $k \longmapsto \widetilde{k}={\Lambda}k$ induces the unitary transformation
$a_{\vec{k}s} \longmapsto U_{\Lambda}a_{\vec{k}s}U_{\Lambda}^{-1}
= \sqrt{E_{\widetilde{\vec{k}}}/E_{\vec{k}}}\;a_{\widetilde{\vec{k}}s}$
(where the spin quantization axis is directed along the boost or rotation axis)~\cite{Peskin:1995}
and thus $|\vec{k},s\rangle\stackrel{\Lambda}{\longmapsto}|\widetilde{\vec{k}},s\rangle$.
From the natural requirement of similarity of transformation for the state \eqref{WPState},
\begin{equation*}
\label{a'}
|\vec{p},x,s\rangle \stackrel{\Lambda}{\longmapsto}|\widetilde{\vec{p}},\widetilde{x},s\rangle
\qquad
(\widetilde{p}={\Lambda}p, \enskip \widetilde{x}={\Lambda}x),
\end{equation*}
it immediately follows that
$\Upphi_{s's}\left(\widetilde{\vec{k}},\widetilde{\vec{p}},\widetilde{x};{{\vec{\sigma}}}\right) = 
 \Upphi_{s's}\left(\vec{k},\vec{p},x ;{{\vec{\sigma}}}\right)$.
For consistency with the standard $S$-matrix scattering theory we must require that the state
\eqref{WPState} turns into the state $|\vec{p},s\rangle$ in the plane-wave (PW) limit,
$|\vec{p},x,s\rangle\stackrel{\text{\tiny PW}}{\longmapsto}|\vec{p},s\rangle$.
Since the parameters ${\sigma_i}$ can always be defined as to approach the PW limit as
$\sigma_i\to0$ ($\forall i$), the correspondence principle can be stated in the following explicit way:
\begin{equation}
\label{Correspondence_Principle_Phi}
\lim_{\vec{\sigma}\to0}\Upphi_{s's}\left(\vec{k},\vec{p},x;\vec{\sigma}\right)
=(2\pi)^32E_{\vec{p}}\delta_{s's}\delta(\vec{k}-\vec{p}).
\end{equation}
Below, our concern is only with the \emph{quasistable} WP states very narrow in the momentum space,
i.e., the parameters $\sigma_i$ are assumed to be sufficiently small.
More formally, we define the quasistable WP as the state whose norm does not depend on the
space-time coordinate $x$ in any inertial reference frame.
Since the right-hand part of Eq.~\eqref{Correspondence_Principle_Phi} is the $x$-independent
relativistic invariant, the correspondence principle suggests that
  (i) function $\Upphi_{s's}$ must be invariant;
 (ii) the $x$-dependence of the function $|\Upphi_{s's}|$ is of no significance;
 \footnote{In fact, the $x$-dependent contributions may enter into the function $|\Phi_{s's}|$ as
           a series in positive powers of $\sigma_i$ and $(k-p)x$ which are small in comparison with the main
           $x$-independent contribution in the vicinity of the maximum $\vec{k}=\vec{p}$. So the simplest
           choice which does not contradict the correspondence principle is to neglect these contributions.
          }
(iii) $|\Upphi_{s's}|\ll|\Upphi_{ss}|$ for $s'{\ne}s$.
These constraints can be accumulated in the simple ansatz
\begin{equation*}
\label{Ansatz}
{\Upphi_{s's}\left(\vec{k},\vec{p},x;\vec{\sigma}\right)
=\delta_{s's}e^{i\zeta(\vec{k},\vec{p},x)}\phi(\vec{k},\vec{p};\vec{\sigma})},
\end{equation*}
in which ${\phi(\vec{k},\vec{p};\vec{\sigma})}$ is a spin- and coordinate-independent
Lorentz-invariant function (henceforth referred to as ``form factor''), such that
\begin{equation}
\label{CP_phi}
{\lim_{\vec{\sigma}\to0}\phi(\vec{k},\vec{p};\vec{\sigma})=(2\pi)^32E_{\vec{p}}\delta(\vec{k}-\vec{p})}
\end{equation}
and $\zeta(\vec{k},\vec{p},x)$ is a real invariant function such that
$\zeta(\vec{k},\vec{p},0)=\zeta(\vec{p},\vec{p},x)=0$.
Since $\zeta$ is a dimensionless function which, by assumption, does not depend on $\vec{\sigma}$,
it can only depend on the scalar product $(k-p)x$ (the dependence on the scalar variable $x^2$
is excluded by both the translation invariance and the corresponding principle).
The simplest choice is $\zeta=(k-p)x$ with the sign uniquely defined by the requirement that the
point $\vec{x}$ be the center of symmetry of the packet in the intrinsic reference frame
(identified by the condition $\vec{p}=0$) in the coordinate representation (see below).
Another, more formal way to explicitly introduce the space-time dependence is
to transform the state $|\vec{p},0,s\rangle$ by the unitary translation operator $\exp[i(\hat{P}x)]$,
where  $\hat{P}=(\hat{P}_0,\hat{\vec{P}})$ is the 4-momentum operator acting on the Fock space,
$\hat{P}_{\mu}|\vec{k},s\rangle=k_{\mu}|\vec{k},s\rangle$. With $\zeta=(k-p)x$ we obtain, as expected,
\[
e^{i(\hat{P}x)}|\vec{p},0,s\rangle = e^{i(px)}|\vec{p},x,s\rangle,
\quad
\langle\vec{p},x,s|\vec{p},x,s\rangle = \langle\vec{p},0,s|\vec{p},0,s\rangle
\]
(cf.\ to Ref.~\refcite{Naumov:2013vea}). Finally, the quasistable QFT wave packet can be written as
\begin{equation}
\label{WavePacketState}
{|\vec{p},x,s\rangle = \int\frac{d\vec{k}\,e^{i(k-p)x}}
{(2\pi)^32E_{\vec{k}}}\phi(\vec{k},\vec{p})|\vec{k},s\rangle}.
\end{equation}
Here and below the argument ${\vec{\sigma}}$ is omitted for short, but is implied.

Let us reveal some more specific features of the form factor $\phi(\vec{k},\vec{p})$.
We will assume that this function has a sharp peak at the point $\vec{k}=\vec{p}$, whose form is
governed by the parameter set ${\vec{\sigma}}$. Therefore the momentum  $\vec{p}$
is just the most probable momentum of the packet and the form factor $\phi(\vec{k},\vec{p})$
represents, up-to a multiplier, a ``smeared'' $\delta$-function.

The norm of the WP state in the momentum basis is 
\begin{equation*}
\label{<aa>_final}
\langle\vec{p},x,s|\vec{p},x,s\rangle = \int\frac{d\vec{k}|\phi(\vec{k},\vec{p})|^2}{(2\pi)^32E_{\vec{k}}}
                                      = \int\frac{d\vec{k}|\phi(\vec{k},\vec{0})|^2}{(2\pi)^32E_{\vec{k}}}.
\end{equation*}
The norm is finite (and momentum independent) if all the parameters $\sigma_i$ are finite,
but it tends to infinity in the PW limit ($\sigma_i\to0$), as it should be, because the normalization
of the Fock states is singular,
$\langle\vec{p}',s'|\vec{p},s\rangle = (2\pi)^22E_{\vec{p}}\delta_{s's}\delta(\vec{p}'-\vec{p})$.
The conventional per-unit normalization would therefore contradict to the correspondence principle.
However, the latter allows us to impose the following Lorentz-invariant condition:
\begin{align}
\label{Normalization_of_phi}
\int\frac{d\vec{k}\,\phi(\vec{k},\vec{p})}{(2\pi)^32E_{\vec{k}}}=
\int\frac{d\vec{k}\,\phi(\vec{k},\vec{0})}{(2\pi)^32E_{\vec{k}}}=1.
\end{align}
Indeed, the dimensionless Lorentz-invariant integral \eqref{Normalization_of_phi} does not depend on $\vec{p}$
and, owing to Eq.~\eqref{CP_phi}, it tends to 1 as $\sigma_i\to0$ ($\forall i$), i.e.\ Eq.~\eqref{Normalization_of_phi}
turns into the identity in the PW limit. It is therefore natural to set condition~\eqref{Normalization_of_phi}
also at finite but small $\sigma_i$.

Treating the wave packet \eqref{WavePacketState} as a physical quantum state produced in collision or decay
of other particles $\varkappa$, one may expect that the form factor depends parametrically
on the ``hidden variables'' -- the most probable 4-momenta $Q_{\varkappa}=(Q_{\varkappa}^0,\vec{Q}_{\varkappa})$
of both primary and secondary packets participated in the production process.%
\footnote{The packet evolution in an external field is also included into this picture since,
          within the $S$-matrix formalism of QFT, any interaction is treated as a local interaction
          of real or virtual fields.}
Moreover, in the most general case the set of the progenitor and accompanying particles may include ones from
the whole net of the reactions and decays which have led to the production of the state \eqref{WavePacketState}.
Being a Lorentz invariant, the form factor can depend on the 4-momenta $k$, $p$ and $Q_{\varkappa}$ only
through the scalar products $(k-p)^2$, $(Q_{\varkappa}k)$, $(Q_{\varkappa}p)$ and $(Q_{\varkappa}Q_{\varkappa'})$.
Owing to the required properties of $\phi$, it is positive-definite in the vicinity $\mathfrak{V}_{\vec{\sigma}}$
of the point $\vec{k}=\vec{p}$ and satisfies the conditions
\begin{align*}
\left[\frac{\partial\phi(\vec{k},\vec{p})}{\partial k_l}\right]_{\vec{k}=\vec{p}}
= &\  \left[\frac{\partial(k-p)^2}{\partial k_l}\,\frac{\partial\phi(\vec{k},\vec{p})}{\partial (k-p)^2}
      \right]_{\vec{k}=\vec{p}}
     +\sum_{\varkappa}\left[\frac{\partial(Q_{\varkappa}k)}{\partial k_l}\,
      \frac{\partial\phi(\vec{k},\vec{p})}{\partial (Q_{\varkappa}k)}\right]_{\vec{k}=\vec{p}} \\
= &\  \sum_{\varkappa}Q_{\varkappa}^0\left(\frac{p_l}{p^0}-\frac{Q_{{\varkappa}l}}{Q_{\varkappa}^0}\right)
      \left[\frac{\partial\phi(\vec{k},\vec{p})}{\partial(Q_{\varkappa}k)}\right]_{\vec{k}=\vec{p}}
=     0 \qquad (l=1,2,3).
\end{align*}
The last equations are satisfied identically only in the unphysical case, when the velocities of all particles
$\varkappa$ are equal to each other ($\vec{Q}_{\varkappa}/Q_{\varkappa}^0=\vec{p}/E_{\vec{p}}$).
Hence, from the arbitrariness of the 4-momentum configurations $\{Q_{\varkappa}\}$ we conclude that
$\left[\partial\phi(\vec{k},\vec{p})/\partial(Q_{\varkappa}k)\right]_{\vec{k}=\vec{p}}=0$ and similarly,
$\left[\partial\phi(\vec{k},\vec{p})/\partial(Q_{\varkappa}p)\right]_{\vec{p}=\vec{k}}=0$. Due to the
analyticity of $\phi$, its dependence upon the scalar products $(Q_{\varkappa}k)$ and $(Q_{\varkappa}p)$
in the small domain $\mathfrak{V}_{\vec{\sigma}}$ must be only through the invariants
\begin{equation}
\label{BuildingBlocks}
g^{\mu\nu}(k-p)_\mu(k-p)_\nu,
\enskip
G_2^{\mu\nu}(k-p)_\mu(k-p)_\nu,
\enskip
G_3^{\mu\nu\lambda}(k-p)_\mu(k-p)_\nu(k-p)_\lambda,
\enskip\ldots,
\end{equation}
where $g$ is the metric tensor and $G_2$, $G_3$, etc.\ are the tensors built from the components
of the 4-vectors $Q_{\varkappa}$ and scalar products $(Q_{\varkappa}Q_{\varkappa'})$. Considering
that the behavior of $\phi$ within the domain $\mathfrak{V}_{\vec{\sigma}}$ is only important,
we conclude that this function has to be constructed from the building blocks \eqref{BuildingBlocks}.
The remaining invariants $(Q_{\varkappa}Q_{\varkappa'})$ can be then ``absorbed'' into the definition
of the parameters $\sigma_i$. In other words, these parameters can be, in general, the scalar functions
of the 4-momenta of the ``network particles'' $\varkappa$ rather than constants.%
\footnote{As a result, the wave packets composed by identical one-particle states but produced
          in different reactions or reaction chains are not, generally speaking, identical.
          Consequently, the quantum statistics for an ensemble of such ``packets with memory'' is expected
          to be quite different from that for their elementary constituents (the states with definite momenta).
         }

As is argued in Ref.~\refcite{Naumov:2010um}, the wave function of the WP state \eqref{WavePacketState}
in the configuration space is given by its projection onto the state $\Psi(x)|0\rangle$, where $\Psi(x)$
is the relevant free field operator. We consider for definiteness a spin-$\frac{1}{2}$ fermion field
\begin{equation*}
\Psi(x) = \int \frac{d\vec{k}}{(2\pi)^3\sqrt{2E_{\vec{k}}}}
\sum_s\left[a_{\vec{k}s}u_s(\vec{k})e^{-ikx}+b_{\vec{k}s}^{\dagger}v_s(\vec{k})e^{ikx}\right].
\end{equation*}
Then the spinor wave function of the packet can be evaluated as
\begin{align}
\langle{0}|\Psi(x)|\vec{p},y,s\rangle
& = e^{-ipy}\left[u_s(\vec{p})-\nabla_{\vec{p}}u_s(\vec{p})
    \cdot\left(i\nabla_{\vec{x}}+\vec{p}\right)+\ldots\right]
    \psi(\vec{p},y-x), \cr
\label{psi_meaning_fermion_c}
& \approx e^{-ipy}u_s(\vec{p})\psi(\vec{p},y-x),
\end{align}
where we have defined the Lorentz-invariant function
\begin{gather}
\begin{aligned}
\label{psi(p,x)}
\psi(\vec{p},x)
= \int\frac{d\vec{k}\,\phi(\vec{k},\vec{p})e^{ikx}      }{(2\pi)^32E_{\vec{k}}}
= \int\frac{d\vec{k}\,\phi(\vec{k},\vec{0})e^{ikx_\star}}{(2\pi)^32E_{\vec{k}}}
= \psi(\vec{0},x_\star),
\end{aligned} \\
\label{x_star_a}
x_{\star}^0=\varGamma_{\vec{p}}\left(x^0-\vec{v}_{\vec{p}}\vec{x}\right),
\quad
\vec{x}_{\star}
=\vec{x}+\varGamma_{\vec{p}}\left[\frac{\varGamma_{\vec{p}}(\vec{v}_{\vec{p}}\vec{x})}
{\varGamma_{\vec{p}}+1}-x^0\right]\vec{v}_{\vec{p}},
\quad
\varGamma_{\vec{p}}=\frac{E_{\vec{p}}}{m},
\quad
\vec{v}_{\vec{p}}=\frac{\vec{p}}{E_{\vec{p}}}. \nonumber
\end{gather}
The spinor wave function $\uppsi_x(\vec{p},y,s)=\langle{0}|\Psi(x)|\vec{p},y,s\rangle$ is the QFT
analog of the QM wave function $\uppsi_{\vec{x}}(\vec{p}_a,x_a)$ and the Lorentz-invariant factor $\psi(\vec{p},y-x)$
in Eq.~\eqref{psi_meaning_fermion_c} defines the nontrivial space-time behaviour of the wave function
with respect to the center of symmetry of the packet $y$.
The function $\psi(\vec{p},x)$ satisfies the Klein–Gordon equation, $(\square-m^2)\psi(\vec{p},x)=0$,
thus representing a relativistic (bosonic) wave packet in terms of the standard scattering theory of QFT
(see, e.g., Ref.~\refcite{Bogolyubov:1990kw}).
The approximation \eqref{psi_meaning_fermion_c} is valid under the condition
\begin{equation}
\label{Condition_for_psi}
\left|i\nabla_{\vec{y}}\ln\psi(\vec{p},x-y)+\vec{p}\right| \ll 2E_{\vec{p}},
\end{equation}
obtained by using the explicit form of the Dirac bispinor $u_s$ (as defined in Ref.~\refcite{Peskin:1995}).
Obviously, this inequality cannot be fulfilled for arbitrary $\vec{p}$ and $x-y$, but is consistent
with the conditions of quasistability used below.
\footnote{Notice that the vector $\vec{\pi}(\vec{p},x-y)=-i\nabla_{\vec{y}}\ln\psi(\vec{p},x-y)$ can be treated
          as a complex-valued function of quantum momentum and one may expect that $\vec{\pi}(\vec{p},x-y)=\vec{p}$
          in the classical trajectories $\vec{x}=\vec{y}+\left(\vec{p}/E_{\vec{p}}\right)\left(x_0-y_0\right)$.
          So, the inequality \eqref{Condition_for_psi} represents the quasiclassicality condition.
          We mention in passing that the equality $\langle{0}|\Phi(y)|\vec{p},0,x\rangle=e^{-ipx}\psi(\vec{p},x-y)$
          is exact for a spinless (scalar or pseudoscalar) field $\Phi(y)$.}
The last equality in Eq.~\eqref{psi(p,x)} is written in the intrinsic reference frame (IRF) of the packet;
hereafter, the variables in this frame are marked by star symbol ${}^\star$ (hence $\vec{p}^\star=0$).
From Eq.~\eqref{psi(p,x)} and the correspondence principle it follows, as expected, that
$\psi(\vec{p},x) \to e^{i(px)}$ as $\sigma\to 0$.

The effective spatial volume of the packet, 
\begin{equation}
\label{EffectiveVolume_QFT}
\mathrm{V}(\vec{p}) \stackrel{\text{def}}{=} \int d\vec{x}|\psi(\vec{p},x)|^2
= \int\frac{d\vec{k}}{(2\pi)^3}\frac{|\phi(\vec{k},\vec{p})|^2}{(2E_{\vec{k}})^2}
= \frac{\mathrm{V}(\vec{0})}{\varGamma_{\vec{p}}}
\end{equation}
is an integral of motion. Therefore the function $|\psi(\vec{p},x)|^2/\mathrm{V}(\vec{p})$ can be treated
as the volume probability density distribution for the state with the most probable momentum $\vec{p}$
in the space-time point $x$. The probability density quickly vanishes as $|\vec{x}|^3\gg\mathrm{V}(\vec{p})$.

\section{Relativistic Gaussian packet}
\label{RGP}

Among the terms \eqref{BuildingBlocks}, only the first one does not depend on the hidden variables.
The simplest relevant model form factor $\phi(\vec{k},\vec{p})$ which satisfies all the conditions
imposed in Sect.~\ref{Wave-packetStates} has been suggested in Ref.~\refcite{Naumov:2010um}:
\begin{equation}
\label{phi_RG_final(b)}
  \phi(\vec{k},\vec{p}) = N_G\exp\left[\frac{(k-p)^2}{4\sigma^2}\right] \equiv \phi_G(\vec{k},\vec{p}).
\end{equation}
Here $\sigma=\text{const}$, $m$ is the field mass and it is assumed that $\sigma^2 \ll m^2$.
The normalization factor $N_G$ in \eqref{phi_RG_final(b)} is defined by the condition \eqref{Normalization_of_phi}
and is equal to $2\pi^{3/2}m/\sigma^{-3}\left[1+3\sigma^2/(4m^2)+O\left(\sigma^4/m^4\right)\right]$.
Although the function \eqref{psi(p,x)} is found in the explicit form\cite{Naumov:2010um},
here we limit ourselves to the asymptotic expansion of its logarithm written in IRF. It can be proved that
\begin{multline}
\label{log_psi_AsymptoticExpansion_star}
\psi(\vec{0},x_{\star})
=  \exp\left\{imx^0_{\star}\left[1+\frac{3\sigma^2}{m^2}
  -\frac{\sigma^4}{m^4}\left(2m^2\vec{x}_{\star}^2-\frac{3}{2}\right)\right] \right. \\
   \left.-\sigma^2\vec{x}_{\star}^2-\frac{3\sigma^4}{m^2}\left[(x^0_{\star})^2+|\vec{x}_{\star}|^2\right]
  +\mathcal{O}\left(\frac{\sigma^6}{m^6}\right)\right\} \equiv \psi_G(\vec{0},x_{\star}).
\end{multline}
An elementary analysis shows that the asymptotic expansion in the exponent of Eq.~\eqref{log_psi_AsymptoticExpansion_star}
can be cut off after the lowest-order terms in $\sigma^2/m^2$ under the following conditions:%
\footnote{For fermionic WPs, the second inequality in Eq.~\eqref{TheRestrictions_psi} has to be replaced by a bit more rigid one,
          $\sigma|\vec{x}_{\star}| \ll m/\sigma$, in order to take into account the important restriction \eqref{Condition_for_psi}.}
\begin{equation}
\label{TheRestrictions_psi}
\sigma^2(x^0_{\star})^2     \ll m^2/\sigma^2,
\quad
\sigma^2|\vec{x}_{\star}|^2 \ll m^2/\sigma^2.
\end{equation}
It is apparent that the space-time region restricted by these inequalities becomes arbitrarily wide as $\sigma\to0$.
Under the conditions \eqref{TheRestrictions_psi}, the function \eqref{log_psi_AsymptoticExpansion_star} becomes
very simple and physically transparent:
\begin{equation}
\label{psi_AsymptoticExpansion_lowest}
\psi_G(\vec{0},x_{\star})
= \exp\left(imx^0_{\star}-\sigma^2|\vec{x}_{\star}|^2\right)
~\stackrel{\Lambda}{=}~
\exp\left\{i(px)-\frac{\sigma^2}{m^2}\left[(px)^2-m^2x^2\right]\right\}.
\end{equation}
We call this function the contracted RGP (CRGP). It is, in particular, seen that
  (i) $\psi_G(\vec{0},x_{\star})$ behaves as a plane wave if $|\vec{x}_{\star}|^2\ll\sigma^{-2}$
      (that is nearby the CRGP center);
 (ii) $|\psi_G(\vec{0},x_{\star})|$ does not depend on the time variable $x^0_{\star}$
      (that is CRGP does not spread);
(iii) $|\psi_G(\vec{0},x_{\star})|$ undergoes Gaussian decay at large distances from the center,
      $|\vec{x}^\star| \gtrsim \sigma^{-1}$;
(iv) $|\psi_G(\vec{p},x)|$ is invariant relative to the group of uniform rectilinear motions
     $\{x_0 \mapsto x_0+\theta, \vec{x}\mapsto\vec{x}+\vec{v}_{\vec{p}}\theta\}$, where
     $|\theta|<\infty$ and $\vec{v}_{\vec{p}}=\vec{p}/E_{\vec{p}}$;
 (v) $|\psi_G(\vec{p},x)|=1$ 
    along the classical world line $\vec{x}=\vec{v}_{\vec{p}}x_0$ but $|\psi_G(\vec{p},x)|<1$ otherwise.

\begin{figure}[htb]
\centering\includegraphics[width=0.99\linewidth]{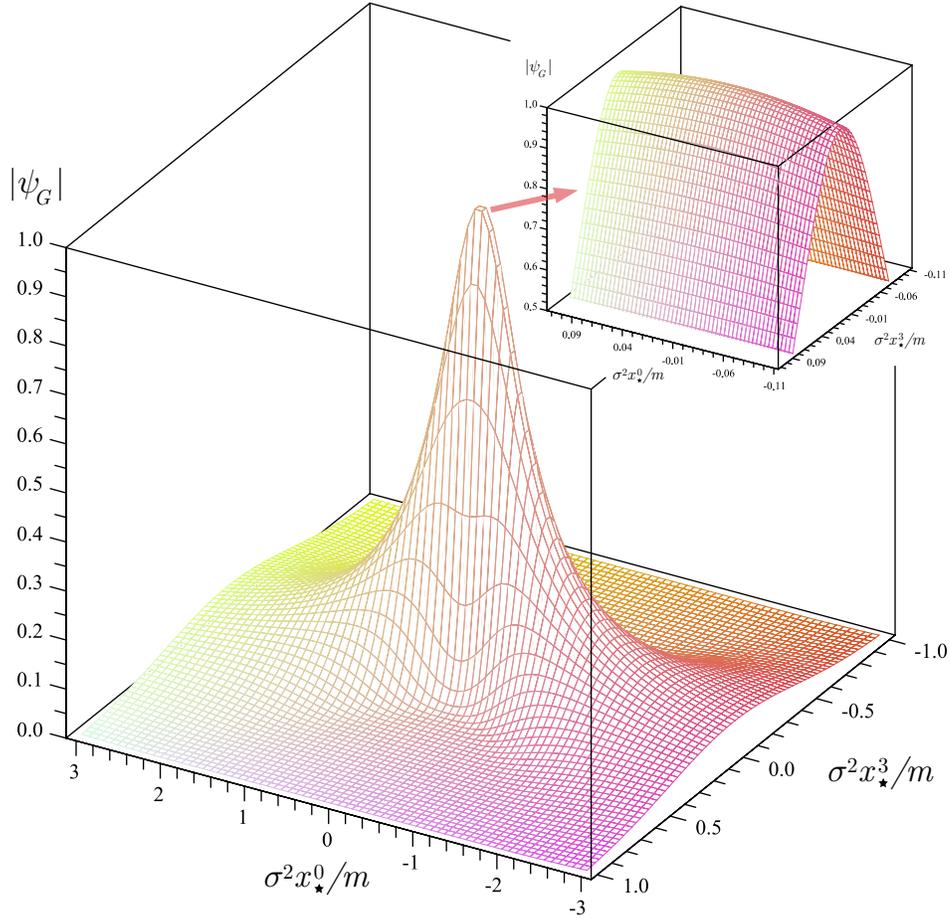}
\caption{\label{fig:mod_psi+zoom} A 3D plot of $|\psi_G(\vec{0},x_\star)|$ as a function of
         $\sigma^2x_\star^0/m$ and $\sigma^2x_\star^3/m$ (vector $\vec{x}_\star$ 
         is directed along the third axis).
         The zoom shows the behaviour of the function $|\psi_G|$ in the space-time region around its peak,
         where the packet is quasistable.
         The calculations are performed with $\sigma/m=0.1$.}
\end{figure}

Figure~\ref{fig:mod_psi+zoom} shows the shape of the \emph{exact} function $|\psi(\vec{p},x)|^2/\mathrm{V}(\vec{p})$
vs.\ dimensionless variables $\sigma^2x_\star^0/m$ and $\sigma^2x_\star^3/m$ (assuming $\vec{x}_\star=(0,0,x_{\star}^3)$)
and its behaviour in the quasistable regime. For better visualization, we use an unrealistic
(in the context of our approximations) ratio $\sigma/m=0.1$
which, however, may be relevant for description of very short-lived hadronic resonances.

According to Eq.~\eqref{EffectiveVolume_QFT}, the effective spatial volume of the CRGP in IRF is
$\mathrm{V}(\vec{0})=(\pi/2)^{3/2}\sigma^{-3}\left[1+O\left(\sigma^2/m^2\right)\right]$.
Since the function $|\psi(\vec{0},x^\star)|^2$ is spherically symmetric and quickly decays at large distances
from the center, the packet can be visualized as a fuzzy ball with the diameter
$d^{\st}=\left[6\mathrm{V}(\vec{0})/\pi\right]^{1/3}=\left(9\pi/2\right)^{1/6}\!\sigma^{-1}$.
So, in the laboratory frame it becomes a spheroid oblate in the direction of motion, owing to the Lorentz contraction.

Let us shortly discuss the conditions of validity of the CRGP model, as applied to description of
the states of unstable particles. To avoid the spreading of the packet during the particle lifetime $\tau$
(that is for $|x^0_{\star}|\lesssim\tau$) we must require that $\sigma^2\tau/m \ll 1$.
Therefore the value $\sigma_{\max}=\sqrt{m/\tau}$ sets the maximum allowable value of $\sigma$
in the CRGP model.
Accordingly, the value $d_{\min}^{\st}=\left(9\pi/2\right)^{1/6}\!/\sigma_{\max}\approx1.6\sqrt{\tau/m}$
sets the minimum allowable value for the effective spatial dimension of the packet in IRF.

As an illustration, Table~\ref{t:Dimension} shows the estimated values of $\sigma_{\max}$ and $d_{\min}^{\st}$
for some long-lived particles. It is seen that in all cases $1/\tau \lll \sigma_{\max} \lll m$.
\begin{table}[htb]
\tbl{Maximum momentum spreads and minimum effective spatial dimensions of CRGP in IRF for several unstable particles.}
{\begin{tabular}{ccccccc}
\hline\noalign{\smallskip} 
Particle     & $\sigma_{\max}$ (eV) & $d_{\min}^\st$ (cm) && Particle    & $\sigma_{\max}$ (eV) & $d_{\min}^\st$ (cm) \\
\noalign{\smallskip}  \hline\noalign{\smallskip}                                                                                                             
$\mu^{\pm}$  & $1.8\times10^{-1}$   & $1.7\times10^{-4}$  && $D^{\pm}$   & $1.1\times10^{ 3}$   & $2.8\times10^{-8}$  \\  
$\tau^{\pm}$ & $2.0\times10^{ 3}$   & $1.3\times10^{-8}$  && $D^{0}$     & $1.7\times10^{ 3}$   & $1.8\times10^{-8}$  \\  
$\pi^{\pm}$  & $1.9$                & $1.6\times10^{-5}$  && $D^{\pm}_s$ & $1.6\times10^{ 3}$   & $1.9\times10^{-8}$  \\  
$\pi^{0}$    & $3.3\times10^{ 4}$   & $0.9\times10^{-9}$  && $B^{\pm}$   & $1.5\times10^{ 3}$   & $2.1\times10^{-8}$  \\  
$K^{\pm}$    & $5.1$                & $6.0\times10^{-6}$  && $B^{0}$     & $1.5\times10^{ 3}$   & $2.0\times10^{-8}$  \\  
$K^{0}_S$    & $6.1\times10^{ 1}$   & $5.1\times10^{-7}$  && $B^{0}_s$   & $1.6\times10^{ 3}$   & $2.0\times10^{-8}$  \\  
$K^{0}_L$    & $2.5$                & $1.2\times10^{-5}$  && $n$         & $2.6\times10^{-5}$   & $1.2$               \\  
\noalign{\smallskip}  \hline\noalign{\smallskip} 
\end{tabular}
\label{t:Dimension}}
\end{table}
The values of $d_{\star}^{\,\min}$ are typically ``mesoscopic''
(i.e., ranging from atomic to millimeter scales), except in the case of the neutron.
But neutrons leave no tracks in the particle detectors... 

\section{Asymmetric wave packet}

Let us now turn to the main topic of the present paper and consider a model
(or, more exactly, a class of models) which accommodates the second term in the list \eqref{BuildingBlocks}.
By analogy with Eq.~\eqref{phi_RG_final(b)}, we introduce the ``truly Gaussian''
Lorentz-invariant form-factor function
\begin{equation}
\label{phi_new}
\phi(\vec{k},\vec{p})=N\exp\left[-\frac{1}{4\sigma^2}\rho_{\mu\nu}(p-k)^\mu(p-k)^\nu\right]
\equiv \phi\strut_{\text{AWP}}(\vec{k},\vec{p}),
\end{equation}
in which $N$ is a normalization constant, $\sigma$ is a small positive parameter 
that defines the scale of the momentum spreading ($\sigma^2 \ll m^2$, $p^2=k^2=m^2$),
and $\rho_{\mu\nu}$ is a symmetric \emph{positive definite} tensor
($\rho_{\mu\nu}q^{\mu}q^{\nu}>0$, $\forall q\ne 0$).
It is apparent that the packet \eqref{phi_new} is not spherically symmetric in the momentum space.
So, we will call it asymmetric wave packet (AWP). 
By applying the standard integral representation for the Dirac $\delta$ function, we can write the
function \eqref{psi(p,x)} as follows:
\begin{equation}
\label{psi_AWP}
\psi(\vec{p},x)=\int dt\int\frac{d^4k}{(2\pi)^42E_{\vec{k}}}\phi\strut_{\text{AWP}}(\vec{k},\vec{p})
                 e^{ikx+it\left(k_0-E_{\vec{k}}\right)}.
\end{equation}
As is proved in Appendix, this integral can be represented in the form
\begin{equation}
\label{FedosovRepresentation}
\psi(\vec{p},x)=\frac{N\sigma^3\exp\left[i(px)-T^{\mu\nu}x_\mu x_\nu\right]}{2\sqrt{\pi^3\omega|\rho|}}D_{\tau}
\exp\left[\frac{W_{\vec{\tau}}}{\omega}\left(i\wrho^{\mu\nu}x_\mu p_{\nu}-\frac{W_{\vec{\tau}}}{4\sigma^2}\right)\right]\Bigg|_{\vec \tau=0}.
\end{equation}
Here $ W_{\vec{\tau}} = E_{\vec{p}}\left(E_{\vec{p}+\vec{\tau}}-E_{\vec{p}}\right)-\vec{p\tau}$,
$\omega=\wrho_{\mu\nu}p^{\mu}p^{\nu}>0$, $\wrho^{\mu\nu}$ is the (positive definite) tensor inverse to 
${\rho}_{\mu\nu}$ ($\wrho^{\mu\lambda}{\rho}_{\lambda\nu}=\delta^{\mu}_\nu$), and $|\rho|=\det||{\rho}_{\mu\nu}||$.
Since the differential operator $D_\tau=\exp\left(\sigma^2\partial_\tau\widetilde{\rho}\partial_\tau\right)$
appeared in the lemma from Appendix acts on the function independent of the zero-component of the 4-vector $\tau$,
it can be written as
\[D_\tau=\exp\left(\sigma^2\partial_{\vec{\tau}}\widetilde{\vec{\rho}}\partial_{\vec{\tau}}\right)
=\exp\left(\sigma^2\wrho_{kn}\frac{\partial}{\partial{\tau_k}}\frac{\partial}{\partial{\tau_n}}\right),
\quad
(k,n=1,2,3).
\]
It is easy to prove that the tensor
\begin{equation}
\label{new_T}
T^{\mu\nu} = \frac{\sigma^2}{\omega}\left(
 \wrho^{\mu\nu    }\wrho^{\lambda\rho}
-\wrho^{\mu\lambda}\wrho^{\nu    \rho}\right)p_{\lambda}p_{\rho}
\end{equation}
is also positive definite. Indeed the invariant quadratic form $\mathfrak{Q}=T^{\mu\nu}x_\mu x_\nu$
can be written in the IRF of the packet where $\mathfrak{Q}$ does not depend on the time variable $x_0^\star$
(since $T^{\st}_{00}=T^{\st}_{0k}=T^{\st}_{k0}=0$): $\mathfrak{Q}=\sigma^2r_{kn}x_k^{\star}x_n^{\star}$,
$r_{kn}=\widetilde{\rho}_{kn}^{\st}-\widetilde{\rho}_{0k}^{\st}\widetilde{\rho}_{0n}^{\st}/\widetilde{\rho}_{00}^{\st}$.

Now, due to the rotation invariance of $\mathfrak{Q}$, 
we direct the $z$-axis along the vector $\vec{x}^{\star}$. Then 
$
\mathfrak{Q} = \left(\sigma^2/\widetilde{\rho}_{00}^{\st}\right)
               \left[\widetilde{\rho}_{00}^{\st}\widetilde{\rho}_{33}^{\st}
                    -\left(\widetilde{\rho}_{03}^{\st}\right)^2\right]|\vec{x}^{\star}|^2 \ge 0.
$
The last inequality is true, because the principal minors of a positive-definite matrix are positive.
The matrix $\vec{r}=||r_{kn}||$ can be therefore diagonalized by an orthogonal transformation
$\vec{x}^\star \mapsto \vec{\chi}=\vec{O}\vec{x}^\star$ and the quadratic form $\mathfrak{Q}$ can be written in terms of the
canonical variables $\chi_k$ ($k=1,2,3$), $\mathfrak{Q} = \sigma^2\sum_kr_k\chi_k^2$,
where $r_k>0$ are the eigenvalues of the matrix $\vec{r}$. 

By representing the differential operator $D_\tau$ as a formal (asymptotic) series in powers of $\sigma^2/m^2 \ll 1$,
Eq.~\eqref{FedosovRepresentation} can be transformed to the following form:
\begin{equation}
\label{AWP_FedosovExpansion}
\psi(\vec{p},x)=
\frac{\sigma^3 N\exp\left[i(px)-T^{\mu\nu}x_{\mu}x_{\nu}\right]}{2\sqrt{\pi^3\omega|\rho|}}
\left[1+\sum\limits_{n=1}^\infty c_n(x)\left(\frac{\sigma}{m}\right)^{2n}\right].
\end{equation}
Since the coefficient functions $c_n(x)$ are invariants, they can be written in terms of variables in IRF of the packet. 
The first two coefficients are 
\begin{align*}
c_1(x^\star) = &\ i\theta_1\mathfrak{s}-\left(\theta_1+\frac{\theta_1^2+2\theta_2}{4\wrho^\st_{00}}\right),                                                        \\
c_2(x^\star) = &\ -\frac{3i}{2}\left(\theta_1^2+2\theta_2
                  +\frac{\theta_1^3+6\theta_1\theta_2+8\theta_3}{6\wrho^\st_{00}}\right)\mathfrak{s}
                  -\frac{1}{2}\left(\theta_1^2+2\theta_2\right)\mathfrak{s}^2                                \\
               &\ +\frac{3}{2}\left(\theta_1^2+2\theta_2+\frac{\theta_1^3+6\theta_1\theta_2+8\theta_3}{3\wrho^\st_{00}}
                  -\frac{\theta_1^4-12\theta_1^2\theta_2-24\theta_4}{16\left(\wrho^\st_{00}\right)^2}\right),
\end{align*}
where 
\[
\mathfrak{s} = \frac{m\wrho^{\st0\mu}x_\mu^\star}{\wrho^\st_{00}},
\quad
\theta_n=\text{Tr}\left(\vec{\wrho}^\st\right)^n=\sum_{k}\varrho_k^n,
\]
and $\varrho_k$ are the eigenvalues of the $3\times3$ matrix $\widetilde{\vec{\rho}}^{\st}=||\wrho_{lm}^{\st}||$.
Taking into account the condition \eqref{Normalization_of_phi} we obtain the equation for determining the
normalization factor $N$:
\[
\frac{\sigma^3 N}{2\sqrt{\pi^3\omega|\rho|}}
\left[1+\sum_{n=0}^\infty c_n(0)\left(\frac{\sigma}{m}\right)^{2n}\right]=1.
\]
Therefore Eq.~\eqref{AWP_FedosovExpansion} may be rewritten as follows (cf.\ to \eqref{log_psi_AsymptoticExpansion_star}):
\begin{align}
\label{psi_asymmetric_after_reexpansion}
\psi(\vec{p},x) 
= &\ \exp\left[i(px)-T^{\mu\nu}x_{\mu}x_{\nu}\right]
      \left\{1+\dfrac{\sum\limits_{n=1}^{\infty}\left[c_n(x)-c_n(0)\right]\left(\sigma/m\right)^{2n}}
                     {1+\sum\limits_{n=1}^{\infty}c_n(0)\left(\sigma/m\right)^{2n}}\right\} \nonumber \\
= &\ \exp\left[i(px)-T^{\mu\nu}x_{\mu}x_{\nu}+\sum\limits_{n=1}^{\infty}\mathfrak{c}_n(x)\left(\frac{\sigma}{m}\right)^{2n}\right],
\end{align}
where the invariant coefficient functions $\mathfrak{c}_n(x)$ are routinely determined through the coefficient functions $c_n(x)$
by division of the series in curly brackets in Eq.~\eqref{psi_asymmetric_after_reexpansion} and subsequent re-expansion of the result.
The two lowest-order coefficients (also written in IRF) are
\begin{equation}
\label{mathfrak_c_i}
\mathfrak{c}_1(x^\star) = i\theta_1\mathfrak{s},
\quad
\mathfrak{c}_2(x^\star) = -\frac{i}{2}\left[\theta_1^2+6\theta_2+\frac{2\left(\theta_1\theta_2+2\theta_3\right)}
                                                   {\wrho^\st_{00}}\right]\mathfrak{s}-\theta_2\mathfrak{s}^2.
\end{equation}
Clearly, at $\sigma^2/m^2 \to 0$ the function $\psi(\vec{p},x)$ behaves as 
\begin{equation}
\label{psi_CAWP}
\exp\left[i(px)-T^{\mu\nu}x_{\mu}x_{\nu}\right] = \exp\left(imx_0^\star-\sigma^2\sum_kr_k\chi_k^2\right)
\end{equation}
and thus the function $|\psi(\vec{0},x^\star)|$ becomes time-independent, in full analogy with
the CRGP case~\eqref{psi_AsymptoticExpansion_lowest}; 
from here on the approximation~\eqref{psi_CAWP} will be referred to as contracted AWP (CAWP).
It can be proved that the CAWP satisfies exactly the same properties (i)--(v) as the CRGP
\footnote{It is pertinent to note that the invariance of the CAWP relative to the group of uniform rectilinear motions
          [property (iv) in Sect.~\ref{RGP}] is a consequence of the identity $T^{\mu\nu}p_{\nu}=0$,
          meaning that the tensor $T$ is transverse to the direction of motion of the packet. The same is also true
          for the CRGP, considering that $T_{\text{CRGP}}^{\mu\nu}=\sigma^2\left(p^{\mu}p^{\nu}/m^2-g^{\mu\nu}\right)$.
          }
with only minor reservations that the conditions of validity of the CAWP approximation should be refined to match
the relations \eqref{psi_asymmetric_after_reexpansion} and \eqref{mathfrak_c_i}. The analog of the conditions
\eqref{TheRestrictions_psi} can be written as follows:
\begin{equation}
\label{StabilityConditionsForAWP}
|\mathfrak{c}_1(x^\star)| \ll \frac{m^3}{\sigma^2}|x_0^\star|,
\quad
|\text{Im}\,\mathfrak{c}_2(x^\star)| \ll \frac{m^5}{\sigma^4}|x_0^\star|,
\quad
|\text{Re}\,\mathfrak{c}_2(x^\star)| \ll \frac{m^4}{\sigma^2}\sum_kr_k\chi_k^2.
\end{equation}
The analysis of these inequalities, as applied to the WP description of the asymptotically free states of unstable
long-lived particles relevant to production of neutrinos, shows that sufficiently
``narrow'' (in the momentum space) AWP remains quasistable (AWP $\approx$ CAWP) over a time comparable with
or longer than the lifetime of the associated particle, much like the CRGP case discussed in Sect.~\ref{RGP}.
It should be however noted that the conditions \eqref{StabilityConditionsForAWP} are in general more severe
than those for the CRGP, since these involve the tensor components $\wrho^{\st}_{\mu\nu}$ (asymmetries) which
may be quite different in magnitude.
According to Eq.~\eqref{EffectiveVolume_QFT}, the effective spatial volume of the CAWP can be estimated as
\[\mathrm{V}(\vec{0})=(\pi/2)^{3/2}(r_1r_2r_3)^{-1/2}\sigma^{-3}\left[1+O\left(\sigma^2/m^2\right)\right].\]

\section{A simple example: AWP created in a two-particle decay.}
\label{AsimpleExample}

Let us consider an example of the tensor $\rho^{\mu\nu}$ inspired by the duality between the propagator and
wave-packet descriptions of the neutrino production/detection process discussed in Ref.~\refcite{Naumov:2013bea}.
In this duality formulation, the effective WP of neutrino is constructed from the so-called inverse overlap tensors
(IOT) of the in and out wave packets involved into the source and detector vertices of the corresponding macroscopic
Feynman diagram. The explicit form of the IOT has been derived within the CRGP model (see Appendix~2 of
Ref.~\refcite{Naumov:2013bea}) and can be used to build up an abstract AWP (not uniquely for neutrinos).
The simplest particular case of IOT,
\begin{equation}
\label{rho_example}
\rho_s^{\mu\nu} = \frac{{\xi}u_1^{\mu}u_1^{\nu}+\xi^{-1}u_2^{\mu}u_2^{\nu}+(u_1u_2)\left(u_1^{\mu}u_2^{\nu}
                 +u_2^{\mu}u_1^{\nu}\right)}{(u_1u_2)^2-1}-g^{\mu\nu},
\end{equation}
involves two 4-velocities $u_1$ and $u_2$ of the in and out packets in the source vertex ($u_1^2=u_2^2=1$)
and corresponds to an AWP production in the decay $1 \to 2+\text{AWP}$. The 4-vectors $u_1$ and $u_2$
(or, more precisely, the independent 3-vectors $\vec{u}_1$ and $\vec{u}_2$) play the role of the ``hidden variables''
mentioned in Sect.~\ref{Wave-packetStates}.
The tensor \eqref{rho_example} depends of the dimensionless positive parameter $\xi$. In the CRGP model,
$\xi=\sigma_1^2/\sigma_2^2$, where $\sigma_1$ and $\sigma_2$ are the momentum spreads of the external packets,
but in our context $\xi$ can be thought as just a parameter which sets the scale of the AWP asymmetries. 

To ensure that the tensor \eqref{rho_example} is positive definite it is sufficient to prove that
$\rho_s^{\mu\nu}q_{\mu}q_{\nu}>0$ for any nonzero 4-vector $q$. Another, a bit more simple way is to study
the eigenvalues of the matrix $\rho_s=||\rho_s^{\mu\nu}||$. The characteristic equation reads
\begin{gather*}
\det(\rho_s-\lambda)=(\lambda-1)\left\{\xi\left[(u_1u_2)^2-1\right]\lambda^3
+\left[\xi+1-4\xi(u_1u_2)u_{10}u_{20}+\xi(u_1u_2)^2 \right. \right. \\
\left. \left. -\xi^2\left(2u_{10}^2-1\right)-2u_{20}^2\right]\lambda^2
+2(\xi+1)\left({\xi}u_{10}^2+u_{20}^2\right)\lambda
-(\xi+1)^2 \right\}=0.
\end{gather*}
To simplify the solution of this equation it is useful to transform the latter into the reference frame where $\vec{u}_1=0$.
Let $|\vec{u}_2|=\upsilon$ in that frame. Then, after simple algebra the above equation can be rewritten as
\begin{equation*}
(\lambda-1)^2\left\{\xi\upsilon^2\lambda^2
-(\xi+1)^2\left[\left(\xi+1+2\upsilon^2\right)\lambda-1\right]\right\}=0,
\end{equation*}
and hence, the sought eigenvalues are
\[
\lambda_1=\lambda_2=1, \quad
\lambda_{3,4}= \left(1+\frac{1}{\xi}\right)
\left[1+\frac{\xi+1}{2\upsilon^2}\pm\sqrt{1+\left(\frac{\xi+1}{2\upsilon^2}\right)^2+\frac{1}{\upsilon^2}}\,\right].
\]
It is easy to check that $\lambda_3>2$ and $0<\lambda_4<1$ as $\upsilon>0$ and $\xi>0$.
The tensor \eqref{rho_example} is therefore positive definite, thus providing the simplest AWP model.

\begin{figure}[htb]
\centering\includegraphics[width=0.99\linewidth]{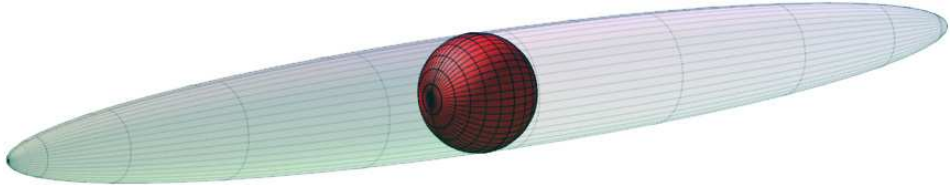}
\caption{\label{fig:ellipsoid} Artist's view of a CAWP spheroid and CRGP ball having the same momentum spread.}
\end{figure}

The tensor inverse to \eqref{rho_example} is given by
\[
\widetilde{\rho}_s^{\mu\nu} =  \frac{\xi}{\xi+1}u_1^{\mu}u_1^{\nu}
                              +\frac{ 1 }{\xi+1}u_2^{\mu}u_2^{\nu}-g^{\mu\nu}.
\]
Its positive definiteness is obvious. In this model, it is easy to find the eigenvalues of the matrix $\vec{r}$
which define the spatial shape of the packet in its IRF. These are
\begin{equation*}
r_1 = 1,
\quad
r_{2,3} = \frac{\xi^2\left|\vec{u}_1^\star\right|^2+\left|\vec{u}_2^\star\right|^2
          +2\left[\mathfrak{U}^2-\xi\left(\vec{u}_1^\star\vec{u}_2^\star\right)u_{10}^{\star}u_{20}^{\star}\right]
          \pm \sqrt{\mathfrak{D}}}{2(\xi+1)\left(\xi\left|\vec{u}_1^\star\right|^2+\left|\vec{u}_2^\star\right|^2\right)},
\end{equation*}
where $\mathfrak{U}^2=\left|\vec{u}_1^\star\right|^2\left|\vec{u}_2^\star\right|^2+\left|\vec{u}_1^\star\right|^2+\left|\vec{u}_2^\star\right|^2$
and
\begin{multline*}
\mathfrak{D} = \xi^4\left|\vec{u}_1^\star\right|^4+\left|\vec{u}_2^\star\right|^4
+4\xi\left(u_1^{\star}u_2^{\star}\right)\left(\vec{u}_1^\star\vec{u}_2^\star\right)
\left(\xi^2\left|\vec{u}_1^\star\right|^2+\left|\vec{u}_2^\star\right|^2\right) \\
+4\xi^2\left\{\left|\vec{u}_1^\star\right|^2\left|\vec{u}_2^\star\right|^2
\left[\mathfrak{U}^2-\left(u_{10}^{\star}u_{20}^{\star}\right)^2+\left(u_1^{\star}u_2^{\star}\right)^2
+\frac{1}{2}\right]+\left(\vec{u}_1^\star\vec{u}_2^\star\right)^2\right\}.
\end{multline*}
In the framework of the diagrammatic approach, the principal contribution to the observables is given by
the momentum configurations of the external in and out packets which satisfy the energy-momentum conservation
to the first order in the momentum spreads $\sigma_i$. In the particular model under consideration, the
quantum fluctuations ${\delta}\vec{p}_i$ of the momenta $m_i\vec{u}_i^\star$ over the physical values
$\vec{p}_i^\star$ satisfying the conditions $\vec{p}_1^\star=\vec{p}_2^\star$,
$\sqrt{|\vec{p}_1^\star|^2+m_1^2}-\sqrt{|\vec{p}_2^\star|^2+m_2^2}=m$ and $m_1>m_2+m$ are bound to be small,
namely, $|{\delta}\vec{p}_i| \lesssim\sigma_i\ll m_i$. As a consequence, we obtain:
\[
r_2 = 1+\frac{2\left[|\delta\vec{p}_1||\delta\vec{p}_2|-\left(\delta\vec{p}_1\delta\vec{p}_2\right)\right]}{m_\xi^2-m^2},
\quad
r_3 =  \frac{m^2}{m_\xi^2}
      -\frac{4(\xi+1)m^3{\delta}Q}{{\xi}m_\xi^4},
\]
where only the leading in $\sigma_i/m_i$ terms are retained and where we used the notation 
\begin{gather*}
{\delta}Q  =  \frac{m_1^2\left[(2{\xi}+1)m_2^2+m_1^2-m^2\right]}{M^2\left(m_1^2-m_2^2+m^2\right)}|\delta\vec{p}_1| 
-\frac{m_2^2\left[(\xi+2)m_1^2+\xi\left(m_2^2-m^2\right)\right]}{M^2\left(m_1^2-m_2^2-m^2\right)}|\delta\vec{p}_2|, \\
M^2        = \sqrt{\left[(m_1+m_2)^2-m^2\right]\left[(m_1-m_2)^2-m^2\right]},
\qquad 
m_\xi^2    = (\xi+1)\left(m_1^2/\xi+m_2^2\right).
\end{gather*}
Thus, in the most important limiting case $\delta\vec{p}_i=0$ corresponding to the exact energy and momentum conservation,
$r_1=r_2=1$ and $r_3=m^2/m_\xi^2$. 
Considering that $m/m_\xi \le m/\left(m_1+m_2\right)<1$ ($\forall\xi$), the volume density $|\psi(\vec{0},x^\star)|^2/\mathrm{V}(\vec{0})$
in the quasistable regime is substantially asymmetric and can be visualized (somewhat symbolically) as a fuzzy prolate ellipsoid
of revolution with the density increasing towards the center.
For $m \ll m_1+m_2$, the CAWP becomes similar to a thin double-edged needle.
Such distribution shape is quite distinct from that of the spherical CRGP cloudlet (see Fig.~\ref{fig:ellipsoid}).

The effective spatial volume of the CAWP in IRF, $\mathrm{V}(\vec{0})\simeq (\pi/2)^{3/2}(m_\xi/m)\sigma^{-3}$, is larger than
that of the CRGP with the same value of the momentum spread $\sigma$.

\section{Conclusions}

In this paper we have proposed a class of models for the covariant asymmetric wave packets (AWP)
which is an alternative to the simplest ``relativistic Gaussian packet'' (RGP) \eqref{phi_RG_final(b)}
intensively used in the QFT approaches to neutrino oscillations.
It is shown that RGP is not, under any circumstances, a particular case of the truly Gaussian AWP,
even though the models have very similar properties in the quasistable regime.
The main physical reason of this is that the AWP tensor $\rho^{\mu\nu}$ can be only defined in terms of
``hidden variables'', the momenta of the wave packets participated in the process (or chain of processes)
of the AWP production, while the RGP does not depend on the history of its origin.
On the other hand, the AWP can be built up from several RGPs, as is demonstrated in Sec.~\ref{AsimpleExample}
on a simple example.
In turn, the AWP states can be used as building blocks to construct more complex wave packets
dependent on the invariants mentioned in the list~\eqref{BuildingBlocks}.

\section*{Appendix}

Here we derive the representation \eqref{FedosovRepresentation} by applying lemma~3.1.4 proved in Ref.~\refcite{Fedosov:1996}.

{\bf Lemma.} 
{\em
Let $\varphi(x) = xAx+ax+a_0$ 
be a quadratic function on $\mathbb{R}^n$, where $A$ is a complex symmetric $n \times n$ matrix with
a positive-definite real part,
$a$ and $a_0$ are a complex vector and a complex constant, respectively. Then for any function $f(x)$
the integral $I = \int dx f(x)\exp[-\varphi(x)]$ may be represented in the form
\begin{gather*}
\label{FedosovLemma-3.1.4}
I = \pi^{n/2}|A|^{-1/2}\exp[-\varphi(z_s)]\left.D_{\tau}T_N(\tau_s,\tau)\right|_{\tau=0}+I_N, \\
|I_N| \le \int d\tau\left|f(x)-T_N(\tau_s,\tau)\right|\exp[-\text{Re}\,\varphi(\tau+\tau_s)],
\end{gather*}
where $D_\tau = \exp\left(\frac{1}{4}\partial_{\tau}A^{-1}\partial_{\tau}\right)$,
$T_N(\tau_s,\tau)$ is the $N$-th order Taylor polynomial of $f(x)$ at $x=\tau_s=\text{Re}\,z_s$,
and $z_s\in\mathbb{C}^n$ is a stationary point of $\varphi(z)$.
}

Note that the stationary point $z_s$ (solution of the linear system $\partial\varphi/\partial x=0$) is given by 
$z_s = -(1/2)A^{-1}a$ and thus $\varphi(z_s) = -(1/4)aA^{-1}a+a_0$.
If $f(x)$ is analytic at the point $z_s$ then $I_N\to0$ as $N\to\infty$ and
\begin{equation}
\label{FedosovLemma-lim}
I = \pi^{n/2}|A|^{-1/2}\exp[-\varphi(z_s)]\left.D_{\tau}f(\tau_s+\tau)\right|_{\tau=0}.
\end{equation}

Consider the special case when 
$\varphi(x)=\varphi(x;t)=xAx+(a+ibt)x+a_0+ic_0t $ and $f(x)=f(x;t)=f_0(x)\exp[-ig(x)t]$,
where the functions $f_0(x)$ and $g(x)$ are analytic in the stationary point of $\varphi(x;t)$, $\forall t\in(-\infty,\infty)$
and $b \in \mathbb{R}^n$. Equation~\eqref{FedosovLemma-lim} yields:
\begin{multline*}
I=I(t) = \sqrt{\frac{\pi^n}{|A|}}\exp\left(\frac{1}{4}aA^{-1}a-a_0\right)D_{\tau}f_0(\tau_s+\tau) \\
       \times\left.\exp\left\{-\frac{1}{4}bA^{-1}bt^2+i\left[\frac{1}{2}aA^{-1}b-c_0-g(\tau_s+\tau)\right]t\right\}\right|_{\tau=0},
\end{multline*}
\begin{multline}
\label{FedosovLemma-lim-Int}
\int_{-\infty}^{\infty}dtI(t) 
  =  2\sqrt{\frac{\pi^{n+1}}{|A|bA^{-1}b}}\exp\left[\frac{1}{4}aA^{-1}a-a_0-\frac{1}{bA^{-1}b}\left(\frac{1}{2}aA^{-1}b-c_0\right)^2\right] \\
     \times \left.D_{\tau}f_0(\tau_s+\tau)
 \exp\left\{\frac{g(\tau_s+\tau)}{bA^{-1}b}\left[aA^{-1}b-2c_0-g(\tau_s+\tau)\right]\right\}\right|_{\tau=0}.
\end{multline}
This result can be directly applied to Eq.~\eqref{psi_AWP} since the $4d$ integral in the right-hand part of Eq.~\eqref{psi_AWP}
is an iterated integral on $\mathbb{R}^4$; we only have to take care of the proper arrangement of the Lorentz indices appearing
in the integrand. Finally, by identifying in Eq.~\eqref{FedosovLemma-lim-Int}
$A      :=  \rho/(4\sigma^2)$,
$a      := -\rho_{\mu\nu}p^\nu/(2\sigma^2)-ix_\mu$,
$b      := -p_\mu/E_{\vec{p}}$,
$a_0    :=  \rho_{\mu\nu}p^{\mu}p^{\nu}/(4\sigma^2)$,
$c_0    := m^2/E_{\vec{p}}$, 
$f_0(x) := E_{\vec{p}}/E_{\vec{k}}$,
$g(x)   := E_{\vec{k}}-\left[m^2+(\vec{pk})\right]/E_{\vec{p}}$,
we arrive at Eq.~\eqref{FedosovRepresentation}. 

\section*{Acknowledgments}

We thank D.~V.~Naumov and an anonymous referee for their useful comments.

\bibliography{references}

\clearpage
\section*{Supplement: Proof of positivity of quadratic form $\rho_s^{\mu\nu}q_{\mu}q_{\nu}$ \\
         {\footnotesize (this section is not included in the journal version)}}
\label{FurtherProofOfPositivity}

\newcommand{\mom}{\boldsymbol{\upkappa}}

Another way to ensure that the tensor \eqref{rho_example} is positive definite is to prove that
\begin{equation*}
\rho_s^{\mu\nu}q_{\mu}q_{\nu}=\frac{{\xi}(u_1q)^2+\xi^{-1}(u_2q)^2+2(u_1u_2)(u_1q)(u_2q)}{(u_1u_2)^2-1}-q^2>0
\end{equation*}
for arbitrary 4-vector $q=(q_0,\vec{q})\ne0$, arbitrary velocities $\vec{u}_{1,2}$, and $0<\xi<\infty$. 
Since $(u_1u_2)>1$ (excluding the unphysical case $\vec{u}_1=\vec{u}_2=0$), it is sufficient
to prove positivity of the function
\[
F(\xi) = \left[(u_1u_2)^2-1\right]\rho_s^{\mu\nu}q_{\mu}q_{\nu}.
\]
This function has global minimum at
\[
\xi=\sqrt{\frac{(u_2q)^2}{(u_1q)^2}}=\left|\frac{(u_2q)}{(u_1q)}\right| \equiv \xi_q
\]
and hence
\begin{equation}
\label{F_q}
F(\xi) \ge F(\xi_q) = 2|(u_1q)(u_2q)|+2(u_1u_2)(u_1q)(u_2q)-\left[(u_1u_2)^2-1\right]q^2.
\end{equation}
For any time-like 4-vector $q$ the scalar function $F(\xi_q)$ can be rewritten in terms
of variables (marked with a star) in the reference frame in which $\vec{q}^\star=0$. We obtain
\begin{align*}
F(\xi_q) 
  & = \left\{2u_{10}^{\star}u_{20}^{\star}\left[(u_1u_2)+1\right]-\left[(u_1u_2)^2-1\right]\right\}\left(q_0^{\star}\right)^2 \\
  & = \left[(u_1u_2)+1\right]\left[u_{10}^{\star}u_{20}^{\star}+(\vec{u}_1^{\star}\vec{u}_2^{\star})
      +1\right]\left(q_0^{\star}\right)^2 \ge 0.
\end{align*}
For the space-like $q$  ($q^2<0$), the inequality \eqref{F_q} is obviously true, as $(u_1q)(u_2q)\ge 0$.
Let us consider the less obvious opposite case, $(u_1q)(u_2q)<0$, in which
\begin{equation*}
F(\xi_q) = \left[(u_1u_2)-1\right]\left\{-q^2\left[(u_1u_2)+1\right]+2(u_1q)(u_2q)\right\}.
\end{equation*}
This function can be rewritten in terms of variables (marked with an asterisk) in the reference frame where
$m_1\vec{u}_1^*=-m_2\vec{u}_2^* \equiv \mom$. In this frame
\begin{equation*}
u_{i0}^*  = \frac{\epsilon_i}{m_i}, \quad
(u_iq)    = \frac{\epsilon_iq_0+(-1)^i(\mom\vec{q})}{m_i}, \quad
(u_1u_2)  = \frac{\epsilon_1\epsilon_2+\mom^2}{m_1m_2},
\end{equation*}
 where we denoted $\epsilon_i = \sqrt{\mom^2+m_i^2}$ ($i=1,2$). Therefore
\begin{multline*}
F(\xi_q) =  \frac{z_+}{m_1^2m_2^2z_-}\left\{ \left[q_0z_-+\left(\epsilon_2-\epsilon_1\right)(\mom\vec{q})\right]^2
           +(m_1-m_2)^2\left[\mom^2\vec{q}^2-(\mom\vec{q})^2\right] \right\} \ge 0,
\end{multline*}
where $z_{\pm}=\epsilon_1\epsilon_2\pm\mom^2-m_1m_2\ge0$.
This accomplishes the proof.

\end{document}